\documentclass[12pt]{article}
\usepackage{amstex}

\columnwidth\textwidth

\begin{document}

\newcommand{\be}{\begin{equation}}
\newcommand{\ee}{\end{equation}}
\newcommand{\beann}{\begin{eqnarray*}}
\newcommand{\eeann}{\end{eqnarray*}}
\newcommand{\bea}{\begin{eqnarray}}
\newcommand{\eea}{\end{eqnarray}}
\newcommand{\nn}{\nonumber}
\newcommand{\ben}{\begin{enumerate}}
\newcommand{\een}{\end{enumerate}}
\newtheorem{df}{Definition}
\newtheorem{thm}{Theorem}
\newtheorem{lem}{Lemma}
\newtheorem{prop}{Proposition}
\begin{titlepage}

\noindent
\hspace*{11cm} BUTP-99/29 \\
\vspace*{1cm}
\begin{center}
{\LARGE What simplified models say \\[0.5cm]
about unitarity and gravitational collapse} 

\vspace{2cm}

P. H\'{a}j\'{\i}\v{c}ek \\
Institute for Theoretical Physics \\
University of Bern \\
Sidlerstrasse 5, CH-3012 Bern, Switzerland \\
\vspace*{2cm}

December 1999 \\ \vspace*{1cm}

\nopagebreak[4]

\begin{abstract}
  This paper is an extended version of a talk at the conference Constrained
  Dynamics and Quantum Gravity QG99. It reviews some work on the quantum
  collapse of the spherically symmetric gravitating thin shell of zero rest
  mass. Recent results on Kucha\v{r} decomposition are applied. The
  constructed version of quantum mechanics is unitary, although the shell
  falls under its Schwarzschild radius if its energy is high enough. Rather
  that a permanent black hole, something like a transient black and white hole
  pair seems to be created in such a case.
\end{abstract}

\end{center}

\end{titlepage}

\section{Introduction}
\label{sec:intro}
The phenomenon of gravitational collapse leads to serious problems in the
classical theory of gravity. The structure of the resulting singularity
contradicts the axioms of the theory such as the equivalence principle. Thus,
the proof of the so-called singularity theorems \cite{H-E} constituted a
strong motivation for many researchers to turn to quantum theory of gravity.

A prominent feature of the classical collapse is the existence of horizons;
they appear sooner than the singularity occurs. Such horizons not only imply
that the singularity is inevitable (this is, roughly, the content of the
singulariry theorems). They also seem to prevent any object or
information from leaving the region of collapse and from coming back to the
asymptotic region. Gradually, the horizon, which is a conceptual, ideal surface
in classical spacetime, was more and more considered as a physical object
of its own: the black hole. Indeed, this way of thinking is very successful in
astrophysics \cite{Thorne} as well as in quantum field theory on a fixed
classical background (eg., the Hawking effect \cite{hawk1}).

It is this fashion of thinking that suggests the existence of processes in
Nature, in which a great amount of structure and information becomes
completely lost. In particular, it was conjectured that in quantum theory,
unitarity will fail \cite{hawk2}, and that black holes possess entropy
associated with the loss of information in collapse \cite{beck}.  These ideas
look very plausible in situations for which the classical fixed spacetime is a
good description, or when the WKB expansion around such a spacetime is a valid
approximation.

The WKB approximation is the dominating approach to collapse and black holes
today. This is understandable, because no full-fledged quantum gravity exists
as yet. However, it may also be expedient to look at simplified
models that include a sort of gravitational collapse on one hand and allow for
a construction of a full quantum theory on the other. It seems fair
to say that the suggestions or hypotheses about the nature of 
gravitational collapse derived from such a study are comparable in importance
and level of credibility to those extracted from the classical theory of
collapse and the WKB expansion around it.

It is also important to realize that the problem of gravitational collapse is
a very special one. For its solution, a complete quantum theory of gravitation
may be as little needed as the complete quantum electrodynamics was for the
first calculations of atomic spectra.

Motivated by these ideas, we consider the quantum theory of a spherically
symmetric thin shell and gravitational field. This is, in fact, a quite
popular system. For example, it was used to study the motion of the domain
walls in the early Universe \cite{guth}, of the black-hole evaporation
\cite{K-W}, of quantum black holes \cite{berez}, and many others.

In Refs.\ \cite{haj1}, \cite{H-K-K} and \cite{honnef}, we have studied the
gravitational collapse of such a thin shell in its own gravitational field.
The result was surprisingly analogous to what is known about the $s$-mode of
the Coulomb scattering. More specifically, two aspects of our result were
surprising.  First, for low-mass shells, there were stationary states with
Sommerfeld spectrum and scattering states with shell's wave packets bouncing and
reexpanding. The evolution was unitary. Second, there was an analogue to the
critical charge in relativistic quantum mechanics of atoms.  The role of charge was
played by the rest mass of the shell (not to be confused with its total
energy) and the critical value of the rest mass was about one Planck mass.  As
in the case of relativistic atoms, the shell quantum mechanics broke
down for supercritical ``charges''.

To understand these results was difficult. Even the simple scattering of the
subcritical shells admitted several different interpretations. There were two
problems.  First, the radial coordinate of the shell, which served as the
argument of the wave function, did not possess the status of quantum
observable. The Coulomb-like potential prevented us from constructing a
position operator similar to the Newton-Wigner one. Second, the model was
completely reduced to the physical degrees of freedom, which in this case was
just the radius of the shell.  However, the value of the radius is not as
informative in a black hole spacetime as it is, say, in Minkowski spacetime:
points with the same value of radial coordinate can lie in very different
asymptotically flat sections.  These sections are separated by horizons. The
tools that were at our disposal in Refs.\ \cite{haj1}, \cite{H-K-K} and
\cite{honnef} did not allow us to decide whether the shell created a horizon
and then, consequently, reexpanded behind this horizon into a different
asymptotically flat section, or it did not create any horizon and reexpanded
into the same section from which it collapsed.

In Ref.\ \cite{haj3}, two remedies were proposed. The first is to work with a
null (lightlike) shell. The classical dynamics of such a shell is equivalent
to that of free photons on flat two-dimensional spacetime (the ``charge'' is
zero). For such a system, there is a well-defined position operator
\cite{wightman}. Moreover, it admits a simple description of its asymptotical
states, unlike Coulomb scattering.

The second idea is to use embedding variables and the resulting decomposition
of the action into the true-degree-of-freedom and embedding parts (the
so-called Kucha\v{r} decomposition, for details see \cite{canada}); the
embeddings contain all necessary information about where the shell is.
However, a transformation of a given action into the Kucha\v{r} form is a hard
problem. In \cite{haj2} and \cite{H-K}, a general method of solving this
problem has been described. The transformation has been split into two steps.
First, the restriction of the transformation to the constraint surface is
well-defined if a gauge is chosen and it is obtained in a straightforward
calculation.  Second, an extension of the resulting variables out of the
constraint surface is to be found. Often, this need not be done explicitly,
the mere knowledge that an extension exists is sufficient. A proof of the
existence based on the Darboux-Weinstein theorem is given in \cite{H-K}. An
application of these results to the shell will be reviewed in the present
paper (details of calculations will be published elsewhere).

The plan of the paper is as follows. Sec.~1 describes the solutions of
Einstein equations containing the shell. These solutions are rewritten as a
set of parameter-dependent metric fields on a fixed background manifold. The
parameters distinguish physically different solutions and the background
manifold is defined by a gauge fixing. Sec.~2 starts from a (non-reduced)
Hamiltonian action principle for the spherically symmetric shell and
gravitational field \cite{L-W-F}. This action is then transformed to the
Kucha\v{r} form corresponding to the gauge fixing.

A construction of a quantum mechanics including the position operator is
described in Sec.~4, and its properties are studied in Sec.~5.  It is
formulated as a dynamics of the shell on the background manifold; this enables
straightforward and unique interpretations. It turns out that 1) there is no
singularity, 2) the dynamics is unitary, 3) no black hole is formed, and 4)
the shell can fall under its Schwarzschild radius, if its energy is
sufficiently high. By the way, these results also prove that the WKB
approximation fails. In Sec.~6, we consider the seemingly contradictory claims
that the shell can cross the Schwarzschild radius and still reexpand. The
explanation is that the quantum shell creates either a black or a white hole
apparent horizon according to the state of its motion; the horizon is black if
the shell is contracting and it is white if the shell is expanding. No event
horizon forms.

\section{Einstein dynamics of the shell}
Any spherically symmetric solution of Einstein's equations with a thin null
shell as the source has a simple structure. Inside the shell, the spacetime is
flat; outside the shell it is isometric to a part of the Schwarzschild
spacetime of mass $M$. The two geometries must be stuck together along a
spherically symmetric null hypersurface so that the points with the same
values of the radial coordinate $r$ coincide.

All physically distinct solutions can be labeled by three parameters: $\eta\in
\{-1,+1\}$, distinguishing between the outgoing, $\eta = +1$, and ingoing,
$\eta = -1$, null surfaces; the asymptotic time of the surface, i.e., the
retarded time $u\in (-\infty,\infty)$ for $\eta = +1$, and the advanced time
$v\in (-\infty,\infty)$ for $\eta = -1$; and the mass $M\in (0,\infty)$. An
ingoing shell creates a black-hole event horizon at $r=2\mbox{G}M$ and ends
up in the singularity at $r=0$. The outgoing shell starts from the singularity
at $r=0$ and emerges from a white-hole particle horizon at $r=2\mbox{G}M$.

To write down a metric of such a solution in an explicit form, a particular
gauge must be chosen. This amounts to fixing a unique coordinate system
$\{x^\mu\}$ in each solution that depends smoothly on the parameters $\eta$,
$u$, $v$, and $M$.\footnote{Gauge fixing can be defined in a coordinate
  independent way, see \cite{haj1} and \cite{H-K}.} The result is a line
element
\[
  ds^2 = g_{\mu\nu}(\eta,u,v,M;x)dx^\mu dx^\nu,
\]
called Kucha\v{r}-Romano-Varadarajan (KRV) metric \cite{K-R-V}. This object
can be considered as a family of metric fields on a fixed {\em background
  manifold} whose points are defined by the coordinates $x^\mu$. The trick
that conjures up a set of metric fields on a fixed manifold out of a set of
spacetimes is the gauge fixing. Different gauges define the points of the
background manifold differently. The background manifold is a {\em naked}
manifold with no other than topological and differential structure: it is
{\em not} a background spacetime with a fixed metric.

A convenient choice of gauge for the shell problem is represented by the
coordinates $U$, $V$, $\vartheta$ and $\varphi$, where $U$ and $V$ are
double-null coordinates, that is, the KRV metric takes on the form
\begin{equation}
  ds^2  =  -AdUdV + R^2d\Omega^2.
\label{KRV}
\end{equation}
The coordinates are completely determined by the following boundary
conditions: 1) At the regular center inside the shell, $U=V$, 2) at the shell,
the functions $U$ and $V$ are continuous, and 3) for the outgoing shells, $U$
is the retarded time for the observers near ${\mathcal I}^+$, for the ingoing
shells, $V$ is the advanced time for the observers near ${\mathcal I}^-$.

To write down the corresponding functions $A$ and $R$ is a straightforward
matter; they are complicated functions of the form
\begin{equation}
  A = A(\eta,u,v,M;U,V),\quad R = R(\eta,u,v,M;U,V). 
\label{A} 
\end{equation}
The trajectory of the shell on the background manifold is simply $U=u$ for
$\eta = +1$ and $V=v$ for $\eta = -1$.

We denote the background manifold by $\mathcal M$. The definition of $\mathcal
M$ is finished by specifying the range of the coordinates $U$ and $V$:
\[
  \frac{-U+V}{2} \in (0,\infty),\quad \frac{U+V}{2}\in(-\infty,\infty).
\]
Thus, there is room for all classical solutions. Although they become singular
at inside points of $\mathcal M$, this does not lead to any difficulty,
because the singularity is a purely classical problem (at least for our
system). In fact, the requirement that all classical solutions can be written
as fields on the background manifold determines the above ranges uniquely.
One may ask how a definition of a background manifold based on the set of
classical solutions is justified. On one hand, if we were to abandon this
requirement, then we would have no guidance whatsoever in choosing the
background manifold.  On the other, well-known quantum mechanical systems can
tunnel only through points of configuration space that are accessible to at
least some classical solutions.

A key property of the background manifold is that it possesses a unique
asymptotic region with ${\mathcal I}^-$ defined by $U \rightarrow -\infty$ and
${\mathcal I}^+$ by $V \rightarrow +\infty$. As the shell cannot run away from
the background manifold, its reappearance at {\em an} asymptotic region must
be interpreted as the reappearance at the asymptotic region of $\mathcal M$.
In this way, the background manifold is a tool to solve the problem of where
the shell reappears. We assume that such a use of the otherwise non-gauge
invariant object $\mathcal M$ does not violate gauge invariance. This is
plausible, because each background manifold must have a unique asymptotic
region in our case, independent of gauge choice.

\section{Canonical formalism}
In this section, a recently published canonical formalism will be transformed
to a form suitable for quantization of the system.

As a Hamiltonian action principle that implies the dynamics of our system, we
take the Louko-Whiting-Friedman (LWF) action (Eq.\ (2.6) of ref.\ 
\cite{L-W-F}). Let us briefly summarize the relevant LWF formulas. The
spherically symmetric metric is written in the form:
\[
  ds^2 = -N^2d\tau^2 + \Lambda^2(d\rho + N^rd\tau)^2 + R^2d\Omega^2,
\]
and the shell is described by its radial coordinate $\rho = {\mathbf
  r}$. The LWF action reads
\[
  S_0 = \int d\tau\left[{\mathbf p}\dot{\mathbf r} + \int
  d\rho(P_\Lambda\dot{\Lambda} + P_R\dot{R} - H_0)\right],
\]
and the LWF Hamiltonian is
\[
  H_0 = N{\mathcal H} + N^\rho{\mathcal H}_\rho,
\] 
where $\mathcal H$ and ${\mathcal H}_\rho$ are the constraints,
\begin{eqnarray}
 {\mathcal H} & = & \frac{\Lambda P_\Lambda^2}{2R} -
 \frac{P_\Lambda P_R}{R} + 
 \frac{RR''}{\Lambda} - \frac{RR'\Lambda'}{\Lambda^2} 
 + \frac{R^{\prime 2}}{2\Lambda} - \frac{\Lambda}{2} + \frac{\eta{\mathbf
 p}}{\Lambda}\delta(\rho - {\mathbf r}),
\label{LWF-H} \\
  {\mathcal H}_\rho & = & P_RR' - P_\Lambda'\Lambda - {\mathbf
 p}\delta(\rho - {\mathbf r});
\label{LWF-Hr}
\end{eqnarray}
the prime denotes the derivative with respect to $\rho$ and the dot that with
respect to $\tau$.

This action will be transformed in two steps. The first step is a
transformation of the canonical coordinates $\mathbf r$, $\mathbf p$,
$\Lambda$, $P_\Lambda$, $R$ and $P_R$ at the constraint surface $\Gamma$ that
is defined by the constraints (\ref{LWF-H}) and (\ref{LWF-Hr}).  The new
coordinates are $u$ and $M$ for $\eta = +1$, $v$ and $M$ for $\eta = -1$, and
the so-called embeddings. In our case, the embeddings are pairs
$(U(\rho),V(\rho))$ of suitable functions on $(0,\infty)$. They have to
represent smooth spacelike surfaces on the background manifold $\mathcal M$
with respect to the KRV metric. Then, each point at the constraint surface
$\Gamma$ determines a unique value of $u$ or $v$, of $M$ and of
$(U(\rho),V(\rho))$. To calculate the corresponding transformation function,
we are allowed to use the KRV metric, as we are on the constraint surface. For
example, $\Lambda = \sqrt{-AU'V'}$, where $A$ is given by Eq.\ (\ref{A}), etc.

The calculation is straightforward but tedious; a particular care must be
taken of boundary terms in the Liouville form. The result is
\[
  S|_\Gamma = \int d\tau(p_u\dot{u} + p_v\dot{v} - np_up_v),
\]
where $S|_\Gamma$ is the action functional restricted to a natural extension
of the constraint surface $\Gamma$, $n$ is a Lagrange multiplier, $U=u(\tau)$
and $V=v(\tau)$ is the shell trajectory in $\mathcal M$, and $p_u$ and $p_v$
are the momenta conjugate to $u$ and $v$. The new constraint $p_up_v = 0$ is
solved either by $p_v = 0$, resulting in the $\eta = 1$ case, or by $p_u = 0$,
resulting in the $\eta = -1$ case. The total energy $M$ of the shell has the
form $M = -p_u - p_v$, and the radius $r$ of the shell is $r = (-u+v)/2$.
Observe that the constraint-surface part of the action is independent of the
embeddings.

The second step is an extension of the functions $u$, $v$, $p_u$, $p_v$,
$U(\rho)$, $P_U(\rho)$, $V(\rho)$ and $P_V(\rho)$ out of the constraint
surface, where the functions $u$, $v$, $p_u$, $p_v$, $U(\rho)$, and $V(\rho)$
are defined by the above transformation and $P_U(\rho)$, and $P_V(\rho)$ by
$P_U(\rho) = P_V(\rho) = 0$. The extension must satisfy the condition that the
functions form a canonical chart in a neighborhood of $\Gamma$. A proof that
such extension exists in general has been given in Ref.\ \cite{H-K}.

The full action that results has the form of the so-called {\em Kucha\v{r}
  decomposition}
\begin{equation}
  S = \int d\tau\left(p_u\dot{u} + p_v\dot{v} - np_up_v\right)
  + \int d\tau\int_0^\infty d\rho(P_U\dot{U} + P_V\dot{V} - H),
\label{KD}
\end{equation}
where $H = N^UP_U + N^VP_V$; $N^U(\rho)$ and $N^V(\rho)$ are Lagrange
multipliers. 

The variables $u$, $v$, $p_u$ and $p_v$ span the effective phase space of the
shell. They contain all true degrees of freedom of the system. The
corresponding part of the action (\ref{KD}) coincides with the action for
free motion of a zero-rest-mass spherically symmetric shell in flat spacetime.
The phase space has non-trivial boundaries:
\begin{equation}
  p_u \leq 0,\quad p_v \leq 0,\quad \frac{-u+v}{2} \geq 0.
\label{boundar}
\end{equation}

The form (\ref{KD}) of the action somewhat obscures the fact that the
classical dynamics of the shell is incomplete, for the classical dynamics of
free zero-rest-mass shells that it resembles is complete. In fact, it is the
spacetime-geometry part of the classical solution around the shell that
prevents the shell from bouncing at the center. This geometry is hidden in the
dependence of the variables $P_U(\rho)$ and $P_V(\rho)$ on the first and
second fundamental forms, $q_{kl}$ and $K_{kl}$, of the surfaces defined by
the embeddings; we do not know these functions explicitly. Still, the geometry
can be obtained indirectly, either from the KRV metric (\ref{KRV}), or,
equivalently, from the original constraints. The new constraint equations,
$P_U(\rho) = P_V(\rho) = 0$, are mathematically equivalent to the old
constraints, Eqs.\ (\ref{LWF-H}) and (\ref{LWF-Hr}). One can, therefore, use
the old constraints to calculate the geometry from the true degrees of freedom
along the hypersurfaces of some foliation. The equivalence of the two methods
within the classical theory (as well as the fact that two spacetimes obtained
by the second method using different foliations are isometric) follows from
the closure of the algebra of Dirac's constraints \cite{teitel}.

\section{Group quantization}
To quantize the system defined by the action (\ref{KD}), we apply the
so-called group-theoretical quantization method \cite{isham}. There are three
reasons for this choice. First, the method as modified for the generally
covariant systems by Rovelli \cite{rovel} (see also \cite{honnef} and
\cite{H-I}) is based on the algebra of Dirac observables of the system;
dependent degrees of freedom don't influence the definition of Hilbert space.
Second, the group method has, in fact, been invented to cope with restrictions
such as Eq.\ (\ref{boundar}). Finally, the method automatically leads to
self-adjoint operators representing all observables. In particular, a unique
self-adjoint extension of the Hamiltonian will be obtained in this way, and
this is the reason that the dynamics will be unitary.

To begin with, we have to find a complete system of Dirac observables. Let us
choose the functions $p_u$, $p_v$, $up_u$ and $vp_v$. Observe that $u$ alone is
constant only along outgoing shell trajectories ($p_u \neq 0$), and $v$ only
along ingoing ones ($p_v \neq 0$), but $up_u$ and $vp_v$ are always
constant. The only non vanishing Poisson brackets are
\[
  \{ up_u,p_u\} = p_u,\quad \{vp_v,p_v\} = p_v.
\]
This Lie algebra generates a group $G_0$ of symplectic transformations of the
phase space that preserve the boundaries $p_u = 0$ and $p_v=0$. $G_0$ is the
Cartesian product of two copies of the two-dimensional affine group.

Handling the last inequality (\ref{boundar}) is facilitated by the canonical
transformation:
\begin{alignat}{2}
  t & = (u+v)/2, & \qquad r & = (-u+v)/2,
\label{tr} \\
  p_t & = p_u + p_v, & \qquad p_r & = -p_u + p_v.
\label{ptpr}
\end{alignat}
The constraint function then becomes $p_up_v = (p_t^2 - p_r^2)/4$.

The positivity of $r$ is simply due to its role as the radius of the
shell. This suggests the following trick. Let us extend the phase space so
that $r \in (-\infty,+\infty)$ and let us define a symplectic map $I$ on
this extended space by $I(t,r,p_t,p_r) = (t,-r,p_t,-p_r)$. The quotient of the
extended space by $I$ is isomorphic to the original space, and we adopt it as
our phase space.

Clearly, only those functions on the extended space that are invariant with
respect to $I$ will define functions on the quotient. Dirac observables of
this kind are, eg., $p_t$, $p_r^2$, the ``dilation'' $D := tp_t + rp_r = up_u
+ vp_v$ and the square of the ``boost'' $J^2 := (tp_r + rp_t)^2 = (-up_u +
vp_v)^2$.

The quantization procedure consists of three steps. First, one extends the
group $G_0$ to $G := G_0\otimes (\mbox{id},I)$. Second, one finds all unitary
irreducible representations of $G$ that satisfy two conditions: (i) The
representatives $\hat{p}_u$ and $\hat{p}_v$ of $p_u$ and $p_v$ have negative
spectra, and (ii) the Casimir operator $\hat{p}^2_t - \hat{p}^2_r$ has value
zero. Luckily, there is only one such representation. Third, one chooses an
eigenspace of the operator $\hat{I}$ as the physical Hilbert space. There are
only two such eigenspaces (with eigenvalues $\pm 1$), and the two
corresponding representations of the algebra generated by $\hat{p}_t$,
$\hat{p}^2_r$, $\hat{D}$ and $\hat{J}^2$ are equivalent, so it does not matter
which one is chosen.

Skipping all details, we describe the resulting quantum mechanics. The
states are determined by complex functions $\varphi(p)$ on ${\mathbf R}_+$;
the scalar product $(\varphi,\psi)$ is 
\[
  (\varphi,\psi) = \int_0^\infty\frac{dp}{p}\varphi^*(p)\psi(p);
\]
the representatives of the above algebra are
\begin{eqnarray*}
 (\hat{p}_t\varphi)(p) & = & -p\varphi(p), \\
 (\hat{p}_r^2\varphi)(p) & = & p^2\varphi(p), \\
 (\hat{D}\varphi)(p) & = & -ip\frac{\partial \varphi(p)}{\partial p}, \\
 (\hat{J}^2\varphi)(p) & = & -p\frac{\partial \varphi(p)}{\partial p} -
 p^2\frac{\partial^2 \varphi(p)}{\partial p^2}. 
\end{eqnarray*}

The next question is that of time evolution. Time evolution of a generally
covariant system described by Dirac observables may seem self-contradictory or
gauge dependent. Here, we apply the approach that has been worked out in
\cite{honnef} and \cite{evol}.  The operator $-\hat{p}_t$ has the meaning of
the total energy $M$ of the system.  We observe that it is a self-adjoint
operator with a positive spectrum and that it is diagonal in our
representation. The parameter $t$ of the unitary group $\hat{U}(t)$ that is
generated by $-\hat{p}_t$ is easy to interpret: $t$ represents the quantity
that is conjugated to $p_t$ in the classical theory and this is given by Eq.\ 
(\ref{tr}). Hence, $\hat{U}(t)$ describes the evolution of the shell states
between the levels of the function $(U+V)/2$ on $\mathcal M$.

The missing piece of information of where the shell is on $\mathcal M$ is
carried by the quantity $r$ of Eq.\ (\ref{tr}). We try to define the
corresponding position operator in three steps.

First, we observe that $r$ itself is not a Dirac observable, but the boost $J$
is, and that the value of $J$ at the surface $t=0$ coincides with $rp_t$. It
follows that the meaning of the Dirac observable $Jp_t^{-1}$ is the position
at the time $t=0$. This is in a nice correspondence with the Newton-Wigner
construction on one hand, and with the so-called evolving constants of motion
by Rovelli \cite{rov2} on the other.

Second, we try to make $Jp_t^{-1}$ into a symmetric operator on our Hilbert
space. As it is odd with respect to $I$, we have to square it. Let us then
chose the following factor ordering:
\begin{equation}
 \hat{r}^2 := \frac{1}{\sqrt{p}}\hat{J}\frac{1}{p}\hat{J}\frac{1}{\sqrt{p}} =
 -\sqrt{p}\frac{\partial^2}{\partial p^2}\frac{1}{\sqrt{p}}.
\label{rsym}
\end{equation}
Other choices are possible; the above one makes $\hat{r}^2$ essentially a
Laplacian and this simplifies the subsequent mathematics.  Third, we have to
extend the operator $\hat{r}^2$ to a self-adjoint one. The Laplacian on the
half-axis possesses a one-dimensional family of such extensions. We apply the
Newton-Wigner symmetry condition \cite{wightman} to narrow the choice: the
eigenfunctions have to transform properly under the subgroup that leaves the
surface $t=0$ invariant. In our case, this subgroup is the dilation one,
generated by $\hat{D}$. Then, we end up with only two possible sets of
eigenfunctions, from which we choose the following:
\begin{equation}
  \Psi_r(p) := \sqrt{\frac{2p}{\pi}}\sin rp,\quad r \geq 0.
\label{rsa}
\end{equation}
This set is already $\delta$-normalized.

\section{Motion of wave packets}
Let us introduce a family of normalized wave packets on the half-axis by
\[
  \psi_{\kappa\lambda}(p) := \frac{(2\lambda)^{\kappa+1/2}}{\sqrt{(2\kappa)!}}
  p^{\kappa+1/2}e^{-\lambda p},
\]
where $\kappa$ is a positive integer and $\lambda$ is a positive number with
dimension of length. The expected energy, which we denote by
$\bar{M}_{\kappa\lambda}$, of the packet is easily calculated to be
$(\kappa+1/2)/\lambda$. $1/\lambda$ can be viewed as a spacing between the
packet energies and also as a spatial width of the packet; then $\kappa$
determines the available energy levels.

The time evolution of the packet is generated by $-\hat{p}_t$:
\[
  \psi_{\kappa\lambda}(t,p) = \psi_{\kappa\lambda}(p) e^{-ipt}.
\]
Let us calculate the corresponding wave function $\tilde{\psi}$ in the
$r$-representation using the $r$-eigenfunctions (\ref{rsa}). The result is
\[
  \tilde{\psi}_{\kappa\lambda}(t,r) = \frac{1}{\sqrt{2\pi}}
  \frac{\kappa!(2\lambda)^{\kappa+1/2}}{\sqrt{(2\kappa)!}}
  \left[\frac{i}{(\lambda +it +ir)^{\kappa+1}} - \frac{i}{(\lambda +it
  -ir)^{\kappa+1}}\right]. 
\]
It follows immediately that
\[
  \lim_{r\rightarrow 0}|\tilde{\psi}_{\kappa\lambda}(t,0)|^2 = 0.
\]
The scalar product measure for the $r$-re\-pre\-sentation is just $dr$, as
$\Psi_r$'s are normalized, so the probability to find the shell between $r$
and $r+dr$ is simply $|\tilde{\psi}_{\kappa\lambda}(t,r)|^2dr$. Our first
important result is, therefore, that the wave packet starts away from the
center $r=0$ and then stays away from it during the whole evolution. This can
be interpreted as the {\em absence of singularity} in the quantum theory: no
part of the packet is squeezed up to a point, unlike the shell in the
classical theory.

A more tedious calculation is needed to obtain the time dependence
$\bar{r}_{\kappa\lambda}(t)$ of the expected radius of the shell. Shuffling
the integration contours in the complex plane helps to obtain the result:
\[
  \bar{r}_{\kappa\lambda}(t) =
  \frac{1}{2\pi}\frac{(\kappa!)^2(2\lambda)^{2\kappa+1}}{(2\kappa)!} 
  \left[\frac{2}{\kappa}\frac{1}{(\lambda^2+t^2)^\kappa}
  + t\int_{-t}^t \frac{dx}{(x^2 + \lambda^2)^{\kappa+1}} +
  \lambda\int_{-\lambda}^\lambda \frac{dx}{(x^2 + t^2)^{\kappa+1}}\right].
\]
To reveal what this long formula hides, let us calculate the asymptotics as
$t\rightarrow\pm\infty$,
\begin{equation}
  \bar{r}_{\kappa\lambda}(t) \approx |t| + \mbox{O}(t^{-2\kappa}),
\label{rinfty}
\end{equation}
as well as the (minimal) expected radius $\bar{r}_{\kappa\lambda}(0)$ at
$t=0$,
\begin{equation}
  \bar{r}_{\kappa\lambda}(0) =
  \frac{1}{\pi}\frac{2^{2\kappa}(\kappa!)^2}{(2\kappa)!}
  \frac{\kappa+1}{\kappa}\frac{\lambda}{\kappa+1/2} > 0.
\label{r0}
\end{equation}

These formulas have a simple interpretation. Let us put the flat metric
\[
 ds^2 = -dUdV + (1/4)(-U+V)^2 d\Omega^2
\]
on the background manifold $\mathcal M$. Then the asymptotic motion of the
center of our packet can be characterized as follows. The center starts at
${\mathcal I}^-$ moving approximately along the ingoing half $V=v$ of a
light cone. It reappears at ${\mathcal I}^+$ moving along the other (outgoing)
half $U=v$ of the same light cone. (This light cone has its vertex at the
center of symmetry, $U=V$, of the background spacetime.) The asymptotic motion
of the shell resembles the motion of free light rays in flat spacetime: the
scattering time delays of these two dynamical systems coincide.

This result is clearly at variance with the classical idea of black hole
forming in the collapse and preventing anything that falls into it from
reemerging. It is, therefore, natural to ask, if the packet is squeezed enough
so that an important part of it gets under its Schwarzschild radius. We can try
to answer this question by comparing the minimal expected radius
$\bar{r}_{\kappa\lambda}(0)$ with the expected Schwarzschild radius
$\bar{r}_{\kappa\lambda H}$ of the wave packet. The Schwarzschild radius is
given by 
\[
 \bar{r}_{\kappa\lambda H} = 2\mbox{G}\bar{M}_{\kappa\lambda} =
 2\frac{\bar{M}_{\kappa\lambda}}{M_P^2},
\]
where $M_P$ is the Planck energy. Now, the values of $\kappa$ and $\lambda$
for which a large part of the packet gets under its Schwarzschild radius
clearly satisfy the inequality
\[
  \bar{r}_{\kappa\lambda}(0) < \bar{r}_{\kappa\lambda H},
\]
or
\begin{equation}
  (\lambda M_P)^2 < 2\pi\frac{\kappa(\kappa + 1/2)^2}{\kappa + 1}
  \frac{(2\kappa)!}{2^{2\kappa}(\kappa!)^2}.
\label{squee1}
\end{equation}

For reasonably broad packets, we have $\lambda M_P \gg 1$. Then the right-hand
side can be estimated by the Stirling formula:
\[
 2\pi\frac{\kappa(\kappa + 1/2)^2}{\kappa + 1}
  \frac{(2\kappa)!}{2^{2\kappa}(\kappa!)^2} \approx \sqrt{2\pi}\kappa.
\]
Substituting this into the inequality (\ref{squee1}) yields
\begin{equation}
  \bar{M}_{\kappa\lambda} > \frac{\lambda M_P}{\sqrt{2\pi}}M_P,
\label{squee2}
\end{equation}
which implies that the threshold energy for squeezing the packet under its
Schwarz\-schild radius is much larger than the Planck energy. For narrow wave
packets, we have that $\lambda M_P \approx 1$, so the inequality
(\ref{squee1}) is satisfied, and the threshold energy is about one Planck
energy. The inequality (\ref{squee2}) is, therefore, always valid. To
summarize: Reasonably narrow packets can, in principle, get under their
Schwarzschild radius; their energy must be much larger than Planck energy.
Even in such a case, the shell bounces as if completely free.

\section{Grey horizons}
In this section, we try to explain the apparently contradictory result that
the quantum shell can cross its Schwarzschild radius in both directions. The
first possible idea that comes to mind is simply to disregard everything
that our model says about Planck regime. This may be justified, because the
model can hardly be considered as adequate for this regime.  However, the
model {\em is} mathematically consistent, simple and solvable; it must,
therefore, provide some mechanism to make the horizon leaky. We shall study
this mechanism in the hope that it can work in more realistic situations, too.

To begin with, we have to recall that the Schwarzschild radius is the radius
of a non-diverging null hypersurface; anything moving to the future can cross
such a hypersurface only in one direction. The local geometry is that of an
apparent horizon. Whether or not an event horizon forms
depends on the geometry near the singularity \cite{H-contra}. As Einstein's
equations are invariant under time reversal, there are, however, two types of
horizons: those associated with a black hole and those associated with a white
hole. Let us call these horizon themselves {\em black} and {\em white}. The
explanation that follows from the model is that quantum states can contain a
mixture of black and white horizons, and that no event horizon forms. We call
such a mixture a {\em grey horizon}.

The existence of grey horizons can be shown as follows. The position and the
``colour'' of a horizon outside the shell is determined by the spacetime
metric. For our model, this metric is a combination of purely gauge and purely
dependent degrees of freedom, and so it is determined, within the classical
version of the theory, by the physical degrees of freedom through the
constraints. In fact, this may be true in more general cases than is
considered in this paper because the so-called uniqueness theorems \cite{H-E}
say, roughly, that equilibrium states of black holes are completely determined
by the dependent degrees of freedom (for example, the monopole of gravity and
electrodynamics, and the dipole of gravity).

The canonical variables of the shell at a foliation hypersurface $\Sigma$
contain the information about its radius $r$, energy $M$, and its direction of
motion $\eta$. A straightforward calculation using the constraints
(\ref{LWF-H}) and (\ref{LWF-Hr}) does reveal that an apparent horizon lies at
$R=2\mbox{G}M$ if $r<2\mbox{G}M$, its colour is white if $\eta = 1$ and black
if $\eta = -1$. We can express this result by saying that matter always
creates such a horizon outside that cannot block its motion. All that matters
is that the shell can bounce at the singularity (which it cannot within the
classical theory).

These results can be carried over to quantum mechanics after quantities such
as $2\mbox{G}M - r$ and $\eta$ are expressed in terms of the operators
describing the shell. Then we obtain a ``quantum horizon'' with the ``expected
radius'' $2\mbox{G}\bar{M}$ and with the ``expected colour'' to be mostly
black at the time when the expected radius of the shell crosses the horizons
inwards, neutrally grey at the time of the bounce and mostly white when the
shell crosses it outwards.

This proof has, however, three weak points. First, the spacetime metric on the
background manifold is not a gauge invariant quantity; although all gauge
invariant geometrical properties can be extracted from it within the classical
version of the theory, this does not seem to be possible in the quantum theory
\cite{haj2}. Second, calculating the quantum spacetime geometry along
hypersurfaces of a foliation on a given background manifold is foliation
dependent\footnote{Some connection may exist between this fact and the
  well-known anomaly in the Dirac algebra of quantum constraints (an analogous
  anomaly plays a great role in the string theory \cite{G-S-W}).  Indeed, an
  anomaly in this algebra and the foliation dependence of the quantum
  spacetime geometry on a given background manifold are related \cite{teitel},
  \cite{kuch-anom}.}. For example, one can easily imagine two hypersurfaces
$\Sigma$ and $\Sigma'$ belonging to different foliations, that intersect each
other at a sphere outside the shell and such that $\Sigma$ intersects the
shell in its ingoing and $\Sigma'$ in its outgoing state. Third, it does not
seem possible to associate the quantity $\eta$ with any self-adjoint operator
of the quantum theory.

The essence of the first two problems is the gauge dependence of the results
of the calculation. However, it seems that this dependence concerns only
details such as the distribution of different hues of grey along the horizon,
not the qualitative fact that the horizon exists and changes colour from
almost black to almost white. Still, a more reliable method to establish the
existence and properties of grey horizons would require another material
system to be coupled to our model; this could probe the spacetime geometry
around the shell in a gauge-invariant way.

As to the third problem, consider for example the radial momentum $p_r$ of the
shell. The sign of $p_r$ is, in fact equal to $\eta$. However, our quantum
theory contains only the operator $p_r^2$. Another attempt is to calculate the
time derivative of the surface area of the shell by the standard formula
$d\hat{r}^2/dt := -i[\hat{r}^2,\hat{H}]$; one obtains immediately:
\[
  \frac{d\hat{r}^2}{dt} = 2i\frac{d}{dp} - \frac{i}{p}.
\]
This is a symmetric operator (with respect to the Radon-Nykodym measure
$p^{-1}dp$) but it has no self-adjoint extension in our Hilbert space. Indeed,
if it had, then the Hilbert space would contain states with arbitrarily low
negative energy. Hence, we cannot, for example, split a state of the shell
into its ingoing and outgoing components, which would be the clean way to
understand how much of the horizon must be black and how much must be
white. However, we can still use the expectation value of this operator as an
indication of the direction in which the shell is moving, which was how we
obtained the colour mixing.

Even if this result is accepted, it may still seem difficult to imagine any
spacetime that contains an apparent horizon of mixed colours. Nevertheless,
examples of such spacetimes can readily be constructed if the assumption of
differentiability is abandoned. A continuous, piecewise differentiable
spacetime can make sense as a history within the path integral method.

The simplest construction of this kind is based on the existence of an
isometry $\mathcal T$ (that can be considered as time reversal) that maps an
ingoing shell spacetime onto an outgoing one. This can be shown as
follows. The shell spacetime corresponding to the parameter values $\eta = 1$,
$u$ and $M=M_+$ can be covered by retarded Eddington-Finkelstein
coordinates $U$, $R$, $\vartheta$ and $\varphi$ so that the metric has the
form: 
\[
  ds^2 = -\left(1-\frac{2M_+}{R}\right)dU^2 - 2dUdR + R^2d\Omega^2
\]
for $U<u$ and 
\[
  ds^2 = -dU^2 - 2dUdR + R^2d\Omega^2
\] 
for $U>u$. Similarly, the
metric to the parameter values $\eta=-1$, $v$, and $M=M_-$ in advanced
Eddington-Finkelstein coordinates $V$, $R$, $\vartheta$ and $\varphi$ reads:
\[
  ds^2 = -\left(1-\frac{2M_-}{R}\right)dV^2 + 2dVdR + R^2d\Omega^2
\]
for $V>v$ and 
\[
  ds^2 = -dV^2 + 2dVdR + R^2d\Omega^2 
\]
for $V<v$. The map
$\mathcal T$ exists only if $M_+ = M_- = M$, and is then given by 
\[
  {\mathcal T}(U,R,\vartheta,\varphi) = (V,R,\vartheta,\varphi),
\] 
where $V = -U +u+v$.

To start the construction, we choose a spacelike hypersurface $\Sigma_1$
crossing the shell before this hits the singularity in an
($\eta=-1$)-spacetime, and find the corresponding surface ${\mathcal
  T}\Sigma_1$ in an ($\eta=1$)-one.  Then we cut away the part of the
($\eta=-1$)-spacetime that lies in the future of $\Sigma_1$ and the part of
the ($\eta=1$)-one in the past of ${\mathcal T}\Sigma_1$. As these boundaries
are isometric to each other, the remaining halfs can be stuck together in a
continuous way. In the resulting spacetime, the shell contracts from the
infinity until it reaches $\Sigma_1$ at the radius $r_1$; then, it turns its
motion abruptly to expand towards infinity again. There is no
singularity and the spacetime is flat everywhere inside the shell. If $r_1 <
2\mbox{G}M$, then there is a horizon at $R = 2\mbox{G}M$. It comes into
being where the ingoing shell crosses the radius $r = 2\mbox{G}M$ and is
black until it reaches $\Sigma_1$.  Then, it changes its colour abruptly to
white and lasts only until the outgoing shell crosses it again.

The spacelike hypersurface $\Sigma_1$ can be chosen arbitrarily. The
construction can, therefore, be repeated in the future of ${\mathcal
  T}\Sigma_1$ in an analogous way so that we obtain a spacetime with two
``pleats''; the shell contracts, then expands, then contracts again and hits
the singularity. The horizon starts as a black ring, then changes to a white
one, and then it becomes black for all times. This history is, however, not
continuous. Clearly, one can repeat the construction arbitrary many times;
this leads to a ``pleated'' spacetime with a zig-zag motion of the shell and
alternating horizon rings of white and black colour. If the spacetime is to be
singularity free, however, there must be an odd number of pleats and an even
number of rings, beginning with the black ring and ending with a white one.

If there are physical objects that can be called black and white holes,
then our results might be interpreted as showing that the collapse
of the shell can create pairs of black and white holes (see also
\cite{kummer}). This is, however, a non-negligible ``if''.

\subsection*{Acknowledgements}
Helpful discussions with J.~Bi\v{c}\'{a}k, K.~V.~Kucha\v{r} and J.~Whelan are
acknowledged. The work was supported by the Swiss National Fonds and the
Tomalla Foundation Z\"{u}rich.

\end{document}